# Laser-induced disassembly of a graphene single crystal into a nano-crystalline network

*Benjamin Krauss, Timm Lohmann, Dong-Hun Chae, Miroslav Haluska, Klaus von Klitzing, Jurgen H. Smet[*]*

Max-Planck-Institut für Festkörperforschung, Heisenbergstrasse 1, D-70569 Stuttgart, Germany

[*] Corresponding author. E-mail: J.Smet@fkf.mpg.de

We report about investigations of time-dependent structural modifications in single crystal graphene due to laser irradiation even at moderate power levels of 1mW in a diffraction limited spot. The structural modifications have been characterized by in situ scanning confocal Raman spectroscopy, atomic force height microscopy and transport studies. The time evolution of the Raman spectrum reveals two different effects: on a short time scale, dopants, initially present on the flake, are removed. The longer time scale behavior points to a laser induced gradual local decomposition of single crystal graphene into a network of interconnected nano-crystallites with a characteristic length scale of approximately 10 nm due to bond-breaking. The broken bonds offer additional docking sites for adsorbates as confirmed in transport and AFM height studies. These controlled structural modifications may for instance be valuable for enhancing the local reactivity and trimming graphene based gas sensors.

The successful isolation of graphene in 2004 [1] has triggered tremendous research effort. An important impetus has come from the unusual linear dispersion in the band structure of graphene as it promises among others access in table top solid state experiments to phenomena normally reserved for high energy physics. Raman spectroscopy is a versatile and powerful tool to probe elementary excitations such as phonons, excitons, plasmons and magnons. In a confocal arrangement, it is possible to obtain spatial Raman maps with a resolution down to the diffraction limit. Raman spectroscopy has historically played an important role in carbon based systems as it allows to distinguish the different hybridization of bonds (sp, $sp^2$, $sp^3$) and extract information about disorder [2,3]. In layered graphite related systems with $sp^2$-bonding, Raman data can in addition reveal the crystallite size, presence of doping [4] and defects [5]. The existence of a double resonance mechanism even enables studies of the phase breaking length [6] and the distinction between various local arrangements of carbon atoms at the edge such as the zigzag or armchair configurations [7]. Also for graphene studies, Raman has proven a valuable tool as it allows to distinguish monolayer graphene from bilayers and multilayers. While Raman spectroscopy has usually been considered non-invasive, we will show here that laser irradiation of graphene or few-layer graphene either has to be dosed carefully to avoid structural modifications or can on purpose be exploited to induce structural changes in graphene in a controlled manner.

Graphene flakes were prepared on doped Si-substrates covered with a 300 nm thick dry thermal $SiO_2$ using micromechanical cleavage similar to the method described in Ref. 1. Monolayers and bilayers were identified with Raman spectroscopy [8] as well as optical microscopy through a calibration of the black/white contrast as proposed in Ref. 9. The results described here focus on monolayers. Comparable results were obtained on bilayer graphene. The effects become however substantially weaker as the number of layers grows. They are absent in bulk HOPG even at the maximum laser power of 12 mW. The measurements were performed on a combined atomic force microscope (AFM)/scanning confocal Raman setup using wavelengths of 488.0 nm from an argon-ion laser or 632.8 nm from a He-Ne laser and power levels up to 12 mW. The integrated high numerical aperture optics allows a diffraction-limited spot size with a diameter of approximately 400 nm on the sample for 488 nm laser light. The sample is mounted on a 100 μm x 100 μm piezo scanner. Alternatively, the incident laser beam can be deflected by a piezo-driven mirror within an 80 μm x 80 μm range. The reflected light can be analyzed either by a photo multiplier tube or a Raman spectrometer with a peak-to-peak resolution of up to 1 cm$^{-1}$. An AFM-tip can be installed in the vicinity of the laser spot to allow AFM and Raman measurements at the same location.

Figure 1 shows a typical Raman spectrum obtained on a freshly prepared flake with a diffraction-limited 1 mW laser spot of 488 nm wavelength after an acquisition time of 30 s. The following Raman features can be distinguished: The first observable peak appears at ~1590 cm$^{-1}$ and is associated with the zone center in-plane longitudinal optical phonons (middle inset in Figure 1 and Ref. 10). This well-known G peak is characteristic for sp$^2$-hybridized carbon-carbon bonds. The second prominent peak is located at ~2700 cm$^{-1}$. This D* peak originates from a double resonance process [11]. The incoming laser light creates an electron-hole pair and after two inelastic scattering events involving phonons with opposite momenta Raman light is emitted during recombination (right inset in Figure 1). If defects are present, one of the two scattering events can occur elastically (left inset in Figure 1). This yields the defect or D peak. This peak competes with the D* peak and exhibits only half the Raman-shift. The D peak is not observable in our pristine flake, which attests the good crystalline quality and undetectable small concentration of defects. A shape analysis of the D* peak has been successfully used to distinguish single layers from bilayers or multilayers [8]. For monolayer graphene the D* peak can be fitted to a single Lorentzian while the multiple bands in multilayer graphene give rise to a more complex peak structure that requires for instance fitting to 4 Lorentzians for bilayers. Here, the D* peak clearly indicates that the flake is a monolayer. Additional smaller features in the Raman spectrum have been identified as well and were labeled in Figure 1, but are not relevant for the remaining discussions (G* at 3250 cm$^{-1}$, which results from an intravalley double resonance scattering process, and the G+A$_{2U}$ peak at 2460 cm$^{-1}$).

Figure 1 also plots the Raman spectrum (blue line) after the graphene flake has been exposed for 18 hours to the diffraction-limited focused laser spot with an intensity of 1 mW. The Raman spectrum has drastically changed. A strong D peak at 1350 cm$^{-1}$ has emerged. It indicates a large increase in the number of broken sp$^2$ carbon-carbon bonds [5]. Concomitantly a peak has appeared at approximately 2950 cm$^{-1}$ or a Raman shift equal to the sum of the G and D peak Raman shifts. This combined D+G peak is in general only observed in the vicinity of defects. We note also the drop in intensity of the D* peak. The presence of defects also allows the double resonance mechanism with only one inelastic event and the D peak grows at the expense of the D* peak, which involves two inelastic events. The D* peak of the treated flake remains composed of a single Lorentzian despite the apparent local structural modifications of the flake. We point out, that the Raman spectrum obtained after 18 hours of laser exposure is very different from that of amorphous carbon [12], sp$^3$-bonded diamond [13] and graphite oxide [14].

In order to gain more insight into these laser induced modifications, we have performed time-resolved Raman measurements. The laser power was set to a moderate level of 1 mW and the laser spot was focused down to about 500 nm. A Raman spectrum was recorded every 30 seconds. During a time interval of 18 hours spectra were acquired. These spectra were analyzed with respect to the intensity and location of the three most prominent features: the G, D and D* peak. To improve the signal-to-noise ratio, a signal averaging procedure was applied to the extracted data traces. It consisted of a fast Fourier transform (FFT), followed by the removal of high frequency components via a parabolic low pass filter with its maximum at zero frequency and zero transmission beyond the cut-off frequency defined by *1/(mΔt)*. The data was then back transformed. Here, *Δt* refers to the time interval in between two recorded spectra and *m* was set equal to 36. Figure 2a and b display the time dependence of the G peak intensity and position. The intensity initially exhibits a rapid drop. So does the peak position. Both traverse a local minimum after two hours. After five hours the intensity decreases monotonously, while the location of the G peak continues to rise. We will argue that two processes, which occur simultaneously but on different time scales, can account for this behavior. Laser induced heating is held responsible for the removal of dopants. It causes a drop of the G peak intensity and dominates the Raman spectrum initially in the region denoted as regime I in Figure 2b. While dopants continue to be removed, the longer term behavior is mainly attributed to $sp^2$-carbon-carbon bond breaking and the gradual disassembly of the macroscopic single graphene crystal into a network of interconnected graphene nano-crystallites at the location of the laser spot (regime II in Figure 2b). The time evolution of other Raman features also corroborates this interpretation as will be discussed later.

We first consider the influence of doping adsorbates on the Raman spectrum. Without special treatment, our as prepared graphene flakes exhibit the charge neutrality point at positive back-gate voltages on the order of 60 V (see supplementary information S1). Hence, our pristine flakes are doped at the level of about $5 \cdot 10^{12}$ holes/cm$^2$. It has been demonstrated previously through the field effect that the G peak undergoes a red shift as the carrier density drops [15]. In graphene the adiabatic Born-Oppenheimer approximation is not valid. The charge carrier density $n$ enters the electron phonon coupling [4] and causes phonon softening when $n$ decreases. The initial behavior of the G peak, i.e. both the sign and magnitude of the change in the Raman shift, is consistent with a drop in the charge carrier density due to the removal of dopant molecules. Laser heating is presumably responsible for the dopant removal. The behavior is analogous to the influence of heating under vacuum conditions as shown in the supplementary information S1 by combining Raman studies with transport studies in a vacuum chamber. There, it is also shown that a drop in the carrier density causes a red shift of the D* peak as well. This is indeed seen in the experiment of Figure 2. Since the flake is measured under ambient conditions water and oxygen are believed to be among the dominant doping species. The data shown in Figure 2 are generic, however this initial behavior (regime I) is absent, if samples are pre-treated (heat or chemical treatment) to ensure that the charge neutrality point is close to zero back-gate voltage in vacuum (heat treatment) or under ambient conditions (chemical treatment). An example of time dependent Raman data on such a heat treated flake has been included as supplementary information (Figure S2). Contacts were fabricated on these flakes to confirm that the neutrality point before laser treatment was close to zero.

Additional evidence that the initial rapid time dependence of the Raman features is caused by dopant removal comes from the time evolution of the full width at half maximum (FWHM) of the G peak shown in Figure 2d. At time zero the FWHM equals approximately 10.5 cm$^{-1}$. It ascends steeply within the first two hours of laser exposure up to 15.2 cm$^{-1}$. This rise in the FWHM is correlated with the drop of the G peak position. The left inset schematically highlights the case with large p-doping. At high doping levels, the energy of the G phonon (~200 meV) is insufficient to create an electron-hole (e-h)-pair as there are no occupied states available. Consequently, the phonon has a long life time and the Raman peak is expected to be narrow [16]. In the case of low doping (right cone in Figure 2d) the phonon can decay rapidly in an e-h-pair and the FWHM will grow as the charge carrier density drops. This behavior is indeed born out in the experimental data of Figure 2d. For exposure times longer than 2 hours, the FWHM varies more slowly. There is an overall but much slower tendency to increase. Its value remains within a band of 2 cm$^{-1}$ of the value reached after the initial steep ascend. Most of all, the correlation between the behavior of the G peak position and its FWHM expected when variations in the charge carrier density would dominate is absent. The G Peak moves to larger values, which suggests an increase in the carrier density, but this blue shift is neither accompanied by a corresponding drop in the line width nor by a blue shift of the D* peak. Hence, at long exposure times (> 2 hours) other mechanisms predominantly govern the time dependent Raman behavior.

During the first few hours of laser exposure the amplitude of the D peak (regime I, Figure 2e) remains small. It suggests that only few carbon-carbon sp$^2$-bonds are broken. For longer laser beam exposure (regime II in Figure 2e, t > 3-4 hours) however the D peak intensity rises and eventually a substantial amount of sp$^2$-carbon-carbon bonds apparently get cracked. We assert that the single crystal of graphene is gradually disassembled underneath the laser spot and the graphene crystal is converted into a network of interconnected graphene nano-crystallites. Since the photon energy is smaller than the binding energy, two photon processes might be responsible for bond-breaking. The long time scale indicates a low probability for bond-breaking. After 10 hours of laser exposure the intensity of the D peak saturates (regime III in Figure 2b). The bond breaking apparently decelerates or terminates. Tuinstra and co-workers [5] have carried out Raman investigations on pellets composed of single crystals of graphite. The crystallite size was determined from X-ray diffraction. Their studies revealed that the characteristic length scale *d* of the crystallites and the ratio between the D and G peak intensity are in inverse proportions *I(D)/I(G) ~ 1/d*. It simply reflects that the Raman intensity of the D peak is proportional to the percentage of "boundary" in the sample. Based on the reported data, we would conclude from the *I(D)/I(G)* ratio observed in the experiments here (Figure 2e) that the characteristic size of the interconnected nano-crystallites, which form during laser exposure, saturates at an average value of approximately 10 nm. Below it will be shown that this picture of nano-crystallite formation is imposingly confirmed by the behavior of the G peak (Figure 2b).

The increase of the D peak intensity is accompanied by a rise of the G peak position. If due to carbon-carbon bond disruption the graphene flake locally disintegrates into a network of nano-crystalline graphene patches, it is natural to attribute this shift of the G peak to phonon confinement [17]. After 10 hours of irradiation, not only the D peak intensity (Figure 2e) but also the G peak position (Figure 2b) saturates. In the pristine graphene flake, the incident photons only interact with phonons that have essentially zero momentum $q \approx 0$ in order to fulfill momentum conservation [18]. For nano-crystallites with size *d*, the Heisenberg uncertainty principle relaxes this momentum selection rule and also phonon modes with a non-zero momentum up to *Δq ≈ 2π/d* contribute to the Raman intensity [19]. For an average crystallite size of 10 nm, we obtain a maximum momentum transfer *Δq ≈ 0.6/nm*. The Raman active zone center phonon mode exhibits a positive dispersion when moving away from the zone center [20]. Due to the lack of experimental phonon dispersion data for graphene, we resort to reported inelastic X-ray scattering data of graphite to estimate this energy dispersion. A linear approximation yields a slope of *S(LO,Γ) ≈ 13nm/cm*. For 10 nm crystallites, phonon modes with an energy larger by at most 8 cm$^{-1}$ compared with the zone center phonon energy will contribute to the Raman G peak. Since all phonons with *Δq* between 0 and ≈ 0.6/nm take part, the G peak will broaden. The estimated maximum blue shift is fully compatible with the experimental data. We start from a highly p-doped flake with a G peak position close to 1590 cm$^{-1}$ and remove dopants by laser irradiation. The G peak position of undoped graphene is approximately 1583 cm$^{-1}$ [21]. The phonon confinement is expected to cause a blue-shift of less than 8 cm$^{-1}$, so that the G peak should not exceed 1591 cm$^{-1}$ in the limit of long exposure times. This agrees well with the data in Figure 2b. It is purely fortuitous that the initial and final G peak positions are so close. A repeat-experiment on a pre-treated flake, which has the charge neutrality point close to zero as checked by transport data, is discussed in the supplementary information (S2). For such a pre-treated flake the time dependent Raman data is mainly governed by bond-breaking effects and those features attributed previously to dopant removal have largely vanished as the remaining doping adsorbates are minimized. At time t = 0, the G peak lies 1-2 cm$^{-1}$ above where it is expected for an undoped flake. After three hours of exposure it has blue-shifted already by more than 5 cm$^{-1}$ consistent with the above interpretation.

Figure 3 displays spatial maps of the D, G, and D* peaks. The panels on the left are Raman maps for the pristine flake while panels on the right side were recorded after 18 hours of laser exposure. In these experiments, the laser spot was defocused to enlarge the spatial extent of the modified area. The spot size was measured separately. It was equal to 1.5 µm and hence the laser irradiated area can be spatially resolved with confocal Raman spectroscopy. A complete Raman spectrum was recorded for each location (step size 200 nm in both spatial directions) and the intensity was evaluated for the three peaks. Apart from the inevitable blurring due to the diffraction limit, the modifications in the Raman spectrum are indeed spatially confined to the laser spot size. A more accurate estimate of the affected region can be obtained by recording the topography with atomic force microscopy. To minimize the influence of the tip, such measurements were performed in tapping mode. Figure 4a depicts an AFM image of a freshly prepared monolayer. The measured height of the flake is about 1 nm instead of 3.35 Å expected for a monolayer. This discrepancy has been reported previously and is attributed to adsorbed molecules on top of the graphene surface or in between the graphene layer and the substrate [22]. Note that in the tapping mode the AFM "height" may also contain a chemical contrast contribution. After laser exposure the irradiated region shows an additional even higher elevation of approximately 1.5 nm (Figure 4b). The disruption of carbon-carbon bonds offers additional docking sites for adsorbates. We associate this height increase with the adsorption of additional molecules from ambient air when the laser is turned off. This assertion is proved in transport experiments described in the supplementary information (S3). Hence laser irradiation may be used to locally enhance the reactivity of the graphene flake, a property which may be exploited in graphene based sensor [23]. The supplementary information (S3 and S4) describes a systematic study of the height and the charge neutrality point which sets in as a function of the laser exposure time after the laser has been turned off and molecules, which were removed during laser irradiation, are re-adsorbed at the graphene surface exposed to ambient air. It comes as no surprise that both are closely correlated with the time development of the Raman D peak. Finally, we have verified that the observed changes in height only occur on the graphene surface itself. The bare substrate exposed to laser irradiation is essentially left intact, which supports further the $sp^2$-bond disruption picture (supplementary information S5).

In summary, we have demonstrated that laser irradiation of graphene can be invasive and may induce controllable structural modifications. Long laser exposure locally disassembles a single crystalline layer of graphene into a network of interconnected graphene nano-crystallites. Their characteristic size saturates at approximately 10 nm in the limit of long exposure times. The additional "boundaries" provide docking sites for molecular adsorbates, so that the reactivity can be locally enhanced.

Available Supporting Information: Influence of doping on the Raman G and D* peak, details on the time evolution of the D, G and D* peaks of an undoped flake, field effect properties of nano-crystalline graphene, discussion of time resolved AFM measurements and an AFM measurement of laser induced modifications at the edge of a graphene flake. This material is available free of charge via the Internet at http://pubs.acs.org.

FIGURE CAPTIONS

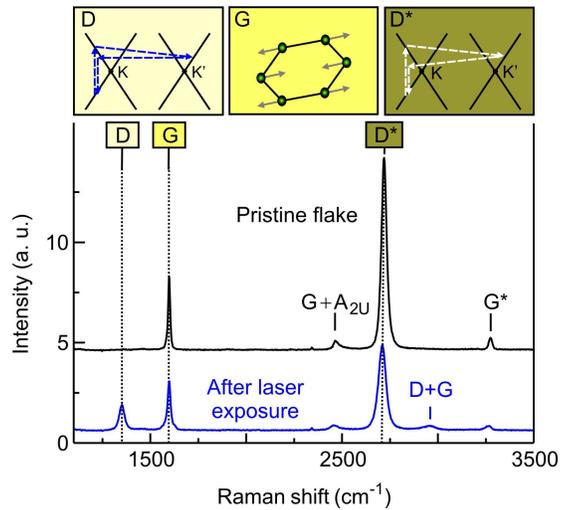

Figure 1. Laser-induced change in the Raman spectrum. The black line (offset in intensity for clarity) is the spectrum of a pristine graphene monolayer. After laser exposure for 18 hours with 1 mW the spectrum dramatically changed (blue line). The appearance of the D peak as well as the decrease in the G and D* peak intensities are a clear signature of the modification.

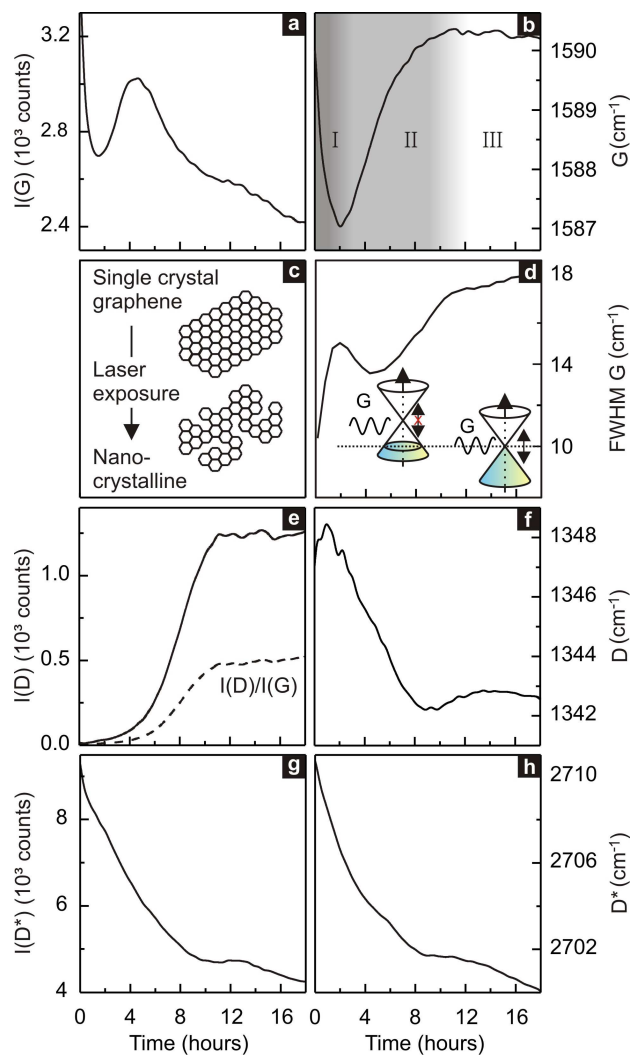

Figure 2. Time evolution of the G, D and D* Raman peaks. Panels a, e and g show the intensity of these peaks as a function of time during laser exposure with 1 mW of power and a wavelength of 488 nm. For each data point on these curves a Raman spectrum was acquired in the range from 1024 cm$^{-1}$ to 3770 cm$^{-1}$ with a 30s integration time. In panel (e) also the intensity ratio between the D and G peaks has been included. Panels b, f and h display the time development of the position of each of these Raman peaks. The FWHM of the G peak is given in (d). The observed changes with time suggest a structural modification of a single crystal of graphene into a network of graphene nano-crystallites (c).

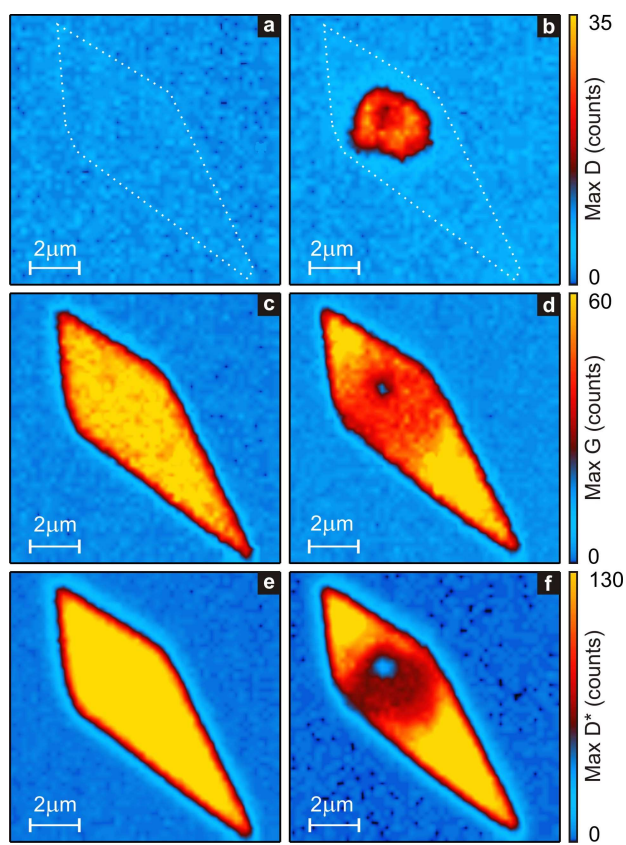

Figure 3. Intensity maps of the D, G, and D* Raman peaks before (left panels) and after laser exposure (right panels). The displayed scale bars are 2 μm. (a) On the pristine flake there are virtually no defects and hence the D Peak is not observable. The white dashed line is a guide to the eye and corresponds to the border of the graphene sheet where a few defects are located. (b) Inside a circle with a diameter of approximately 1.5 μm, the intensity of the D peak is irreversibly enhanced after laser exposure. The size and position of the spot with a large D count coincides with the AFM-image in Figure 4 and the location of the laser spot. (c) Before the laser treatment, the intensity of the G Peak is nearly identical across the entire graphene sheet. It attests to the high crystalline quality of the flake. (d) After laser treatment, the intensity of the G Peak is reduced.

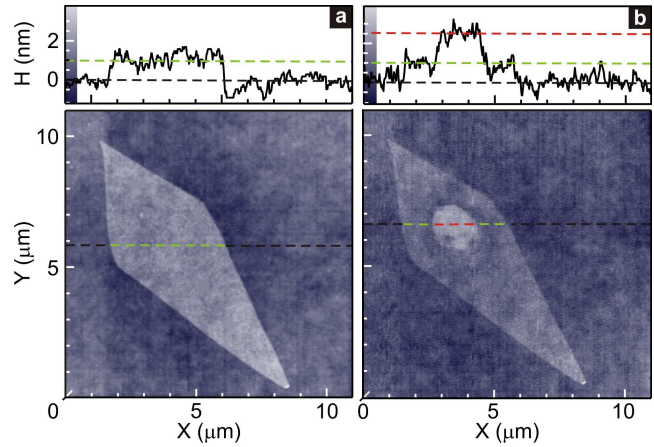

Figure 4. AFM-Image recorded in tapping mode of a graphene monolayer before and after laser exposure. (a) AFM height image of a pristine graphene flake. The top panel shows a height scan along the dashed line. The height (H) of the substrate and the graphene surface are marked by the dashed black and green lines respectively. The measured height difference is approximately 1 nm. (b) AFM height-image after exposing one spot on the sample for 14 hours with a 1mW defocused 488 nm laser beam. An elevated area has appeared where the laser spot was positioned. In the top panel, the red dashed line marks its height. The difference with respect to the pristine graphene surface is approximately 1.5 nm. The diameter of the modified region coincides with the diameter of about 1.5 µm of the defocused laser spot.